\documentclass[twocolumn,epjc3,nospthms]{svjour3}
\usepackage{graphics}
\usepackage[usenames,dvipsnames]{color}
\usepackage{graphicx} 
\usepackage{dcolumn}
\usepackage{bm}
\usepackage{epsfig}
\usepackage{pdflscape}
\usepackage{subfigure} 
\usepackage{multirow}
\usepackage{booktabs}
\usepackage[colorlinks=true,citecolor=blue,filecolor=blue,linkcolor=blue,urlcolor=blue,pdftex]{hyperref}
\usepackage{rotating}
\usepackage{longtable}
\usepackage{lineno}
\usepackage{siunitx}

\begin{document}
\title{First time-resolved measurement of infrared scintillation light in gaseous xenon}

\author{Mona~Piotter\thanksref{addr1}
        \and
        Dominick~Cichon\thanksref{addr1}
        \and
        Robert Hammann\thanksref{e1,addr1}
        \and
        Florian~J\"org\thanksref{addr1}
        \and
        Luisa H\"otzsch\thanksref{addr1}
        \and
        Teresa~Marrod\'an~Undagoitia\thanksref{e2,addr1}
}

\thankstext{e1}{e-mail: robert.hammann@mpi-hd.mpg.de}
\thankstext{e2}{e-mail: teresa.marrodan@mpi-hd.mpg.de}

\institute{Max-Planck-Institut f\"ur Kernphysik, Saupfercheckweg 1, 69117 Heidelberg,
Germany\label{addr1}
}

\date{Received: date / Accepted: date}

\maketitle

\begin{abstract}
\noindent
Xenon is a widely used detector target material due to its excellent scintillation properties in the ultraviolet (UV) spectrum.
The additional use of infrared (IR) scintillation light could improve future detectors.
However, a comprehensive characterization of the IR component is necessary to explore its potential.
We report on the first measurement of the time profile of the IR scintillation response of gaseous xenon.
Our setup consists of a gaseous xenon target irradiated by an alpha particle source and is instrumented with one IR- and two UV-sensitive photomultiplier tubes. Thereby, it enables IR timing measurements with nanosecond resolution and simultaneous measurement of UV and IR signals.
We find that the IR light yield is in the same order of magnitude as the UV yield. We observe that the IR pulses can be described by a fast and a slow component and demonstrate that the size of the slow component decreases with increasing levels of impurities in the gas.
Moreover, we study the IR emission as a function of pressure.
These findings confirm earlier observations and advance our understanding of the IR scintillation response of gaseous xenon, which could have implications for the development of novel xenon-based detectors.
\end{abstract}

\section{Introduction} 
\label{sec:intro}

Xenon is employed as a target medium in different applications including particle-, astroparticle-, neutrino-, and medical physics\,\cite{Chepel:2012sj,Gonzalez-Diaz:2017gxo}. While most detectors use it in its liquid state\,\cite{Auger:2012gs,Abe:2013tc,LZ:2022ufs,XENON:2022kmb,Akimov:2022xvr,Baldini:2016wtm}, there are also experiments utilizing its solid state\,\cite{Kravitz:2022mby} 
or its gaseous state at high pressures\,\cite{Alvarez:2012flf,Han:2017fol}. All these applications use the 175\,nm\,\cite{Jortner:1965s,Fujii_2015xx} ultraviolet (UV) light emitted in the xenon scintillation process.

Infrared (IR) photons are also emitted in the de-excita\-tion process that follows an energy deposition in the xenon me\-dium. This radiation was first detected in the gaseous phase\,\cite{Carungo:1998xx} and, shortly after, in liquid xenon\,\cite{Bressi:2000nim}. 
From these early studies, the infrared signal appeared to be influenced only slightly by the presence of electronegative impurities, while electrons in the charge signal were lost due to oxygen present in the system\,\cite{Bressi:2000nim}. The infrared emission in the gas has been measured to be mainly at a wavelength of $\sim$\SI{1.3}{\micro m}, shifting towards longer wavelengths as the pressure increases\,\cite{Borghesani:2001xx,Borghesani:2008xx}. 
The infrared light yield in gaseous xenon was found to be of a similar magnitude compared to the UV yield\,\cite{Belogurov:2000}, constituting a promising additional signal for future applications. First studies point out that the yield in liquid xenon could be significantly lower than in gas\,\cite{Bressi:2001}. 
Studies of the IR scintillation light of xenon can also be relevant for dual-phase time projection chambers\,\cite{Chepel:2012sj} which record their charge signal via proportional scintillation in the gas phase. The additional detection of IR photons emitted in this process could help, for instance, to improve the energy resolution of this signal, or eventually to improve the discrimination between different types of particles.

In this paper, we report on the observation of infrared photons emitted by gaseous xenon after excitation by alpha particles.
We find a strong dependence of the IR signal on the level of outgassing impurities.
In addition, we analyze the time profile of the IR scintillation response in all measurements. Our observations reveal at least two components with distinct time scales.
Furthermore, we estimate the IR light yield and study the IR signal for gas pressures ranging from 500\,mbar to 1050\,mbar.  

The document is organized as follows: in section\,\ref{sec:xe_scint}, we start introducing the general scintillation mechanism in xenon. Sections\,\ref{sec:exp_set_up} and~\ref{sec:measurements} describe the experimental setup and the measurements performed. Finally, the results are presented and discussed in sections\,\ref{sec:results} and \ref{sec:discussion}.

\section{Xenon scintillation process}   
\label{sec:xe_scint}

\begin{figure}[h]
	\centering
	\includegraphics[width=0.48\textwidth]{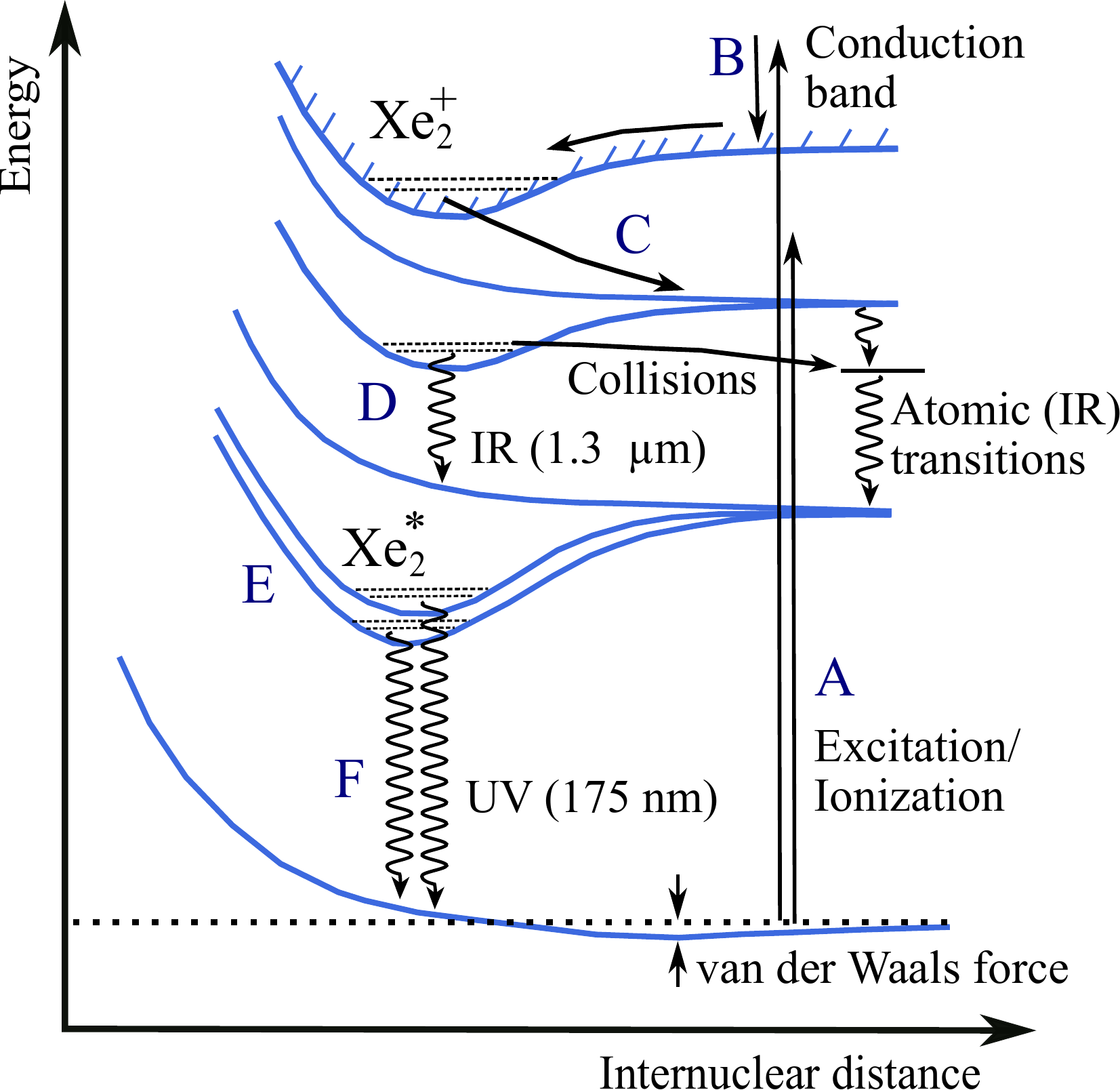}
	\caption{Illustration of excitation and relaxation processes in xenon (not to scale). Infrared scintillation is emitted by atomic or excimer transitions. Figure inspired by~\cite{Borghesani:2001xx} as well as~\cite{Schwentner:1985n,Mulliken:1970rs}}
	\label{fig:Excitons_Scheme}
	\end{figure}

Energy deposited by charged particles in xenon leads to the emission of scintillation light. The target atoms get excited or ionized as shown by process A in figure\,\ref{fig:Excitons_Scheme}. Ionized xenon atoms form Xe$_2^+$ dimers (B). These can recombine with surrounding electrons into excited atomic states (C) which can be also reached by direct excitation. 
Close to atmospheric pressure, the de-excitation of gaseous xenon can happen via different processes like atomic transitions (producing narrow IR lines), via collisions with neighboring atoms, or via an excited dimer (excimer) transition, emitting a broad spectrum centered approximately at \SI{1.3}{\micro m} (D). The measured infrared spectrum of gaseous xenon (redrawn from\,\cite{Borghesani_2007}) is shown later in figure\,\ref{fig:ir_sensitivity} (blue curve) together with the sensitivity of the PMT employed for the measurement. The spectrum contains both the broad excimer emission and a collection of sharp lines from atomic transitions.
Above about 200\,mbar, all excitations end at the lowest lying excimer states Xe$_2^*$ (E). These decay via the emission of ultraviolet photons with a central wavelength of 175\,nm, with two characteristic lifetimes corresponding to a singlet and a triplet transition (labeled F in figure\,\ref{fig:Excitons_Scheme}).

\section{Experimental setup}  
\label{sec:exp_set_up}

\begin{figure}[h]
	\centering
	\includegraphics[width=0.49\textwidth]{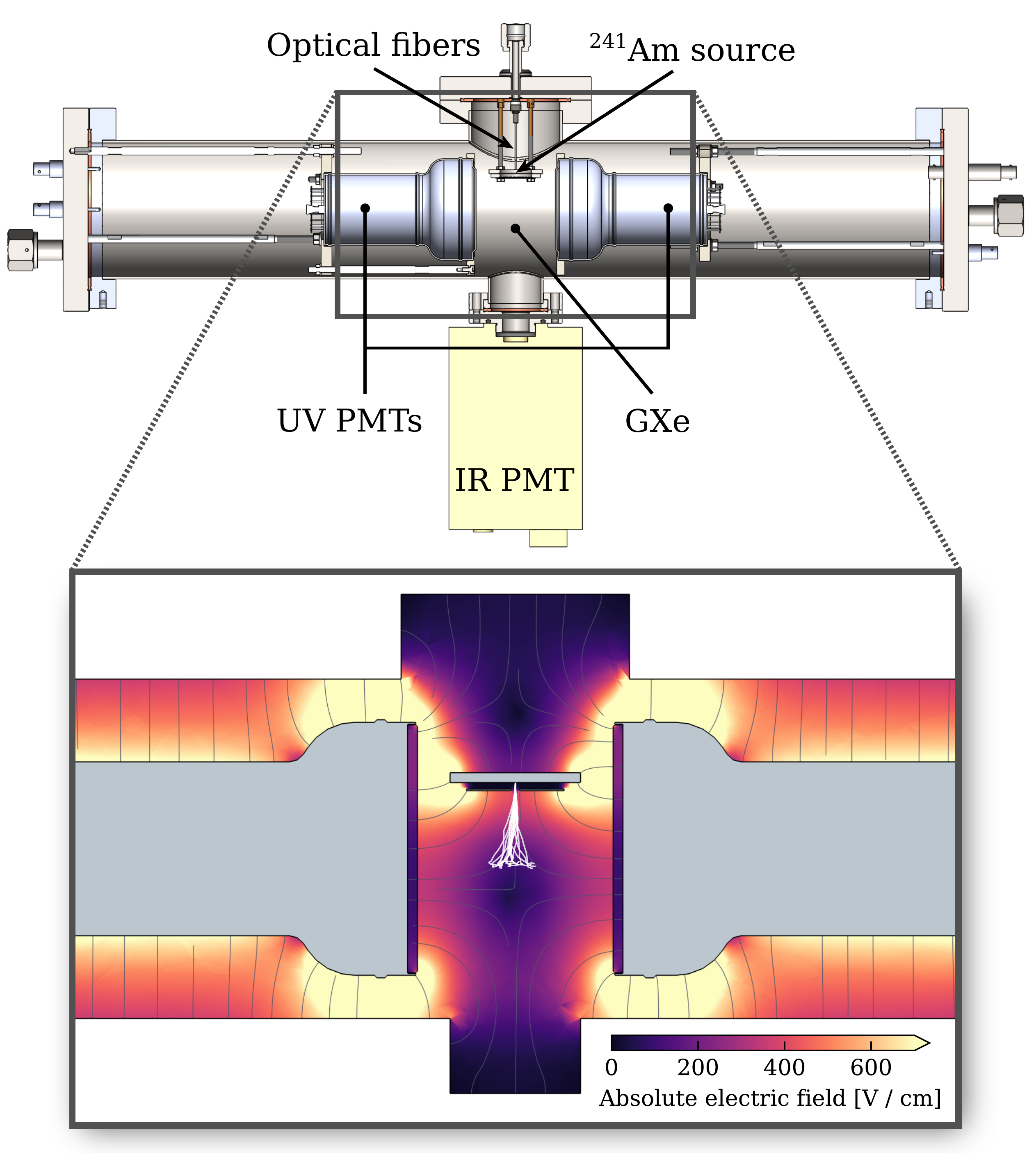}
	\caption{Top: Design of the setup employed to measure infrared radiation in gaseous xenon. Bottom: Electric field simulation (using COMSOL) of the setup's interior. A few typical alpha trajectories (from SRIM) are shown in white}
	\label{fig:ir_setup}
	\end{figure}

The apparatus employed for the measurement consists of a cylindrical chamber of 64\,cm length and 10\,cm diameter (see figure\,\ref{fig:ir_setup}), hosting two UV-sensitive photomultiplier tubes (PMTs), and a connection to an infrared-sensitive PMT (see subsection\,\ref{subsec:photosensors}). 

The chamber volume is filled with gaseous xenon (GXe) and is irradiated with an $^{241}$Am alpha source of 3.7\,kBq activity and an energy of 5.49\,MeV. An aperture with a diameter of 2.5\,mm is placed 3\,mm in front of the source.
In this way, alpha traces are collimated and deposit their energy between the UV PMTs in the gaseous volume.

The signals of all three photomultipliers are recorded and digitized using a CAEN V1743 ADC board, featuring a 3.2\,GS/s effective sampling frequency. To exploit the excellent time resolution of the infrared PMT with narrow pulses of just a few ns, a digitizer with a high sampling frequency is preferred. 
We already employed and characterized this digitizer in\,\cite{Cichon:2022gbu}.

To estimate the magnitude of the electric field in the region where alpha particles deposit their energy, a COMSOL simulation is used\,\cite{comsol:1998}. The simulation is performed in three dimensions and contains all relevant elements of the setup. Figure\,\ref{fig:ir_setup} (bottom) displays the simulated electric field norm in the central plane.
The length of the alpha tracks (in white) is calculated using the SRIM simulation toolkit\,\cite{Ziegler:2010}. The alpha particles travel mostly along the middle axis between the UV PMTs (about 2\,cm for 1\,bar pressure), crossing field regions close to 0 between the source and the aperture, and values from $\sim 400$\,V/cm to about 170\,V/cm behind the aperture.
This electric field caused by the high voltage of the PMTs prevents part of the ionization electrons from recombining.

\subsection{Gas handling system}  
\label{subsec:gas_system}

\begin{figure}[h]
	\centering
	\includegraphics[width=0.48\textwidth]{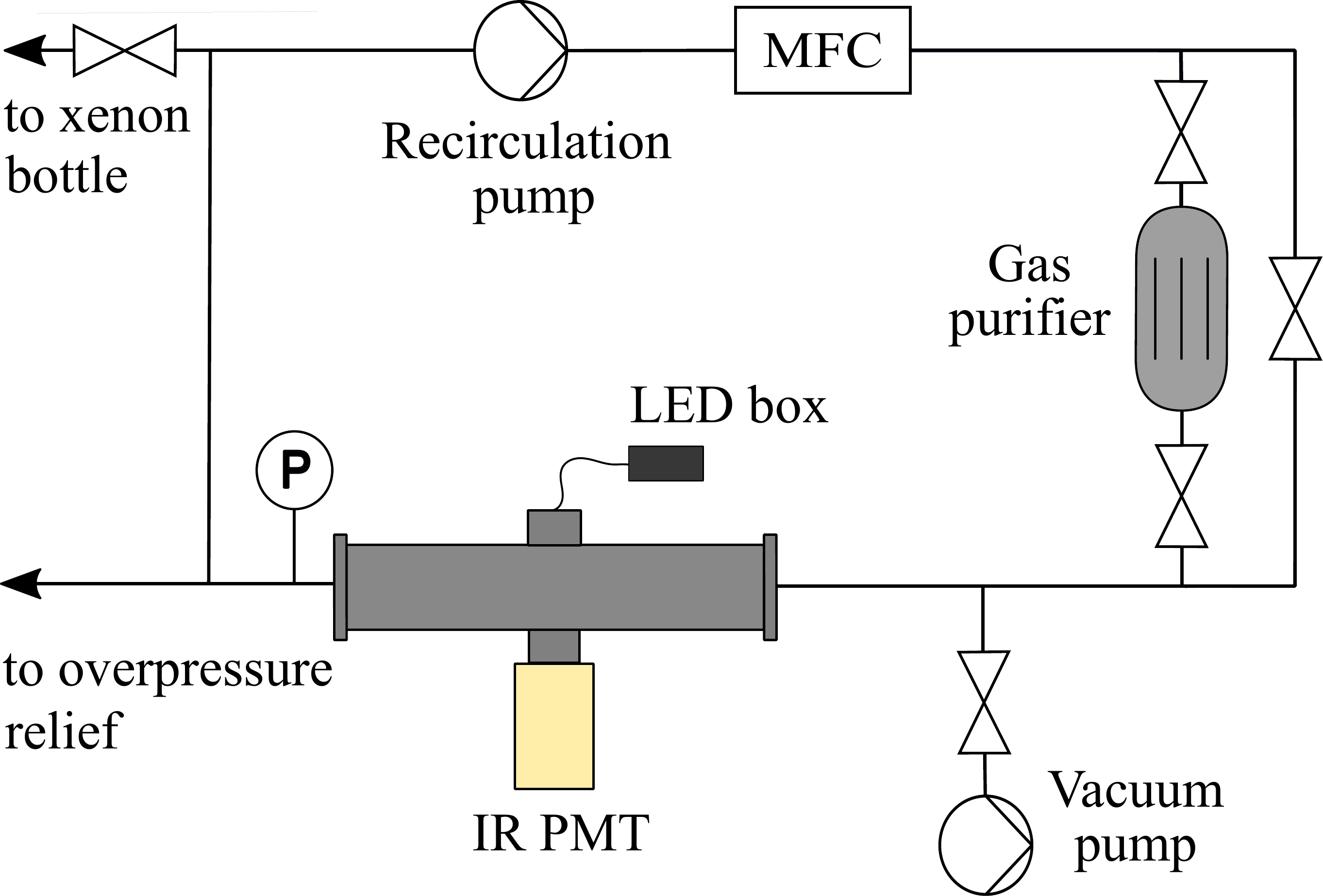} 
	\caption{Schematics of the gas handling system of the setup. Continuous removal of impurities can be achieved by recirculating the gas through a purifier}
	\label{fig:gas_system}
	\end{figure}

To maintain a high level of purity of the gaseous xenon, we use a purification system that continuously removes impurities released from material outgassing. This is important since impurities like O$_2$ or water vapor are known to decrease the UV signal in xenon\,\cite{Ozone2005}. For this purpose, the xenon is recirculated and purified through a zirconium hot getter (SAES MonoTorr PS4). By running the xenon either through the purifier or through its bypass, the effect of the impurities on the UV and IR light yields can be investigated. Figure\,\ref{fig:gas_system} shows a schematic drawing of the purification system.

A small recirculation pump is employed to extract the xenon gas from the chamber and push it through the purifier at a mass flow of $\sim 650$\,sccm at \SI{1000}{mbar}. Throughout a run, the flow is kept at a constant level by a mass flow controller (MFC). The system also includes a vacuum pump, a temperature sensor, and a pressure sensor (P).

\subsection{Response of photosensors}  
\label{subsec:photosensors}

\begin{figure}[h]
	\centering
	\includegraphics[width=0.48\textwidth]{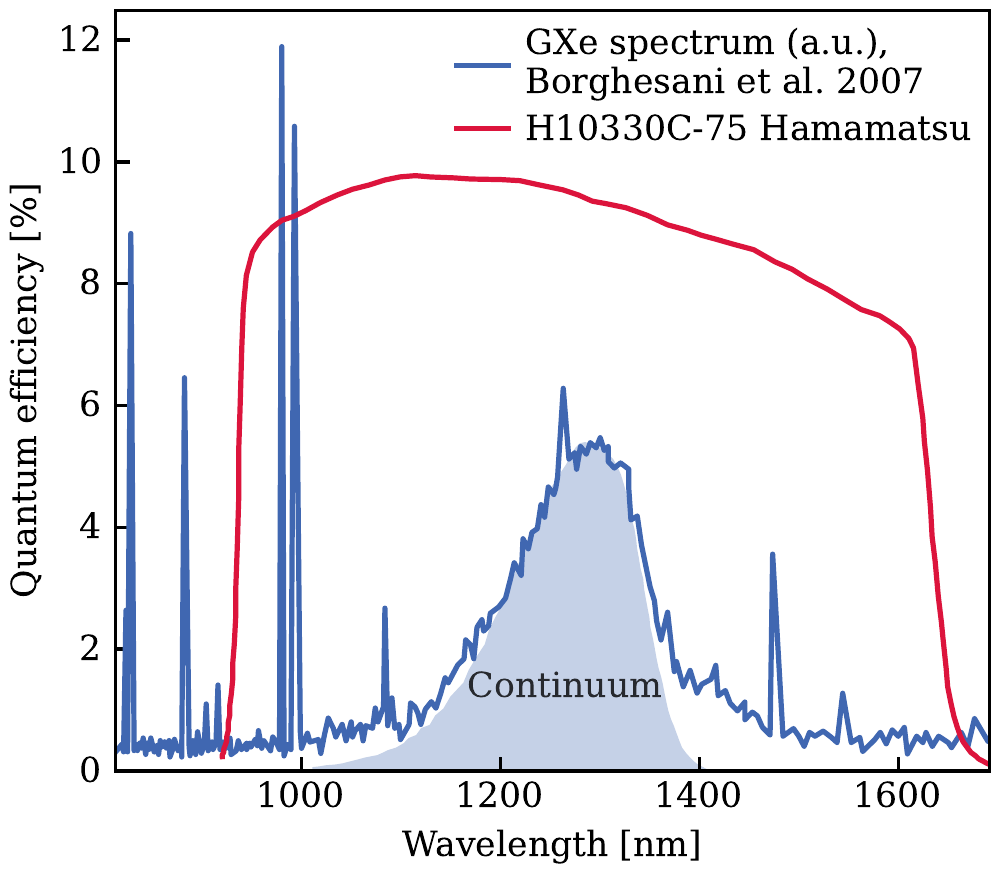}
	\caption{Quantum efficiency (red) of our H10330C PMT (measured by Hamamatsu) together with the infrared emission spectrum of gaseous xenon (blue). Spectral data for 1\,bar xenon redrawn from\,\cite{Borghesani_2007} (arbitrary units). The filled area represents the excimer de-excitation (D in figure\,\ref{fig:Excitons_Scheme})}
	\label{fig:ir_sensitivity}
	\end{figure}

For the measurement of the IR photons, a Hamamatsu H10330C PMT was procured. This PMT has an internal thermoelectric cooler that keeps it at a temperature of about $-60^{\circ}$C. Compared to photodiodes used in previous experiments\,\cite{Belogurov:2000,Bressi:2001}, the IR-PMT has two distinct advantages. Firstly, it is sensitive to single photons, and secondly, it has a fast time response with a rise time and a fall time of 0.9\,ns and 1.7\,ns, respectively.
The fast timing allows studying the time structure of the infrared pulse (see section\,\ref{sec:results}). 
Figure\,\ref{fig:ir_sensitivity} shows the quantum efficiency ($QE$) of our IR-PMT overlaid on top of the gaseous xenon spectrum as measured in~\cite{Borghesani_2007}. 
\label{sec:ir_pmt}
The PMT is sensitive to wavelengths from 950 to $\sim1\,650$\,nm, which includes the excimer transition at \SI{1.3}{\micro m} and several atomic transition lines.

To detect the UV xenon scintillation light, two 3-inch round Hamamatsu R11410 PMTs\,\cite{Aprile:2015lha} are utilized. This sensor type was optimized together with the company, and employed in the XENON1T and XENONnT experiments\,\cite{XENON:2017lvq,Barrow:2016doe,Antochi:2021wik}. Its spectral sensitivity ranges from about 160\,nm to about 600\,nm.

\begin{figure}[h]
	\centering
	\includegraphics[width=0.45\textwidth]{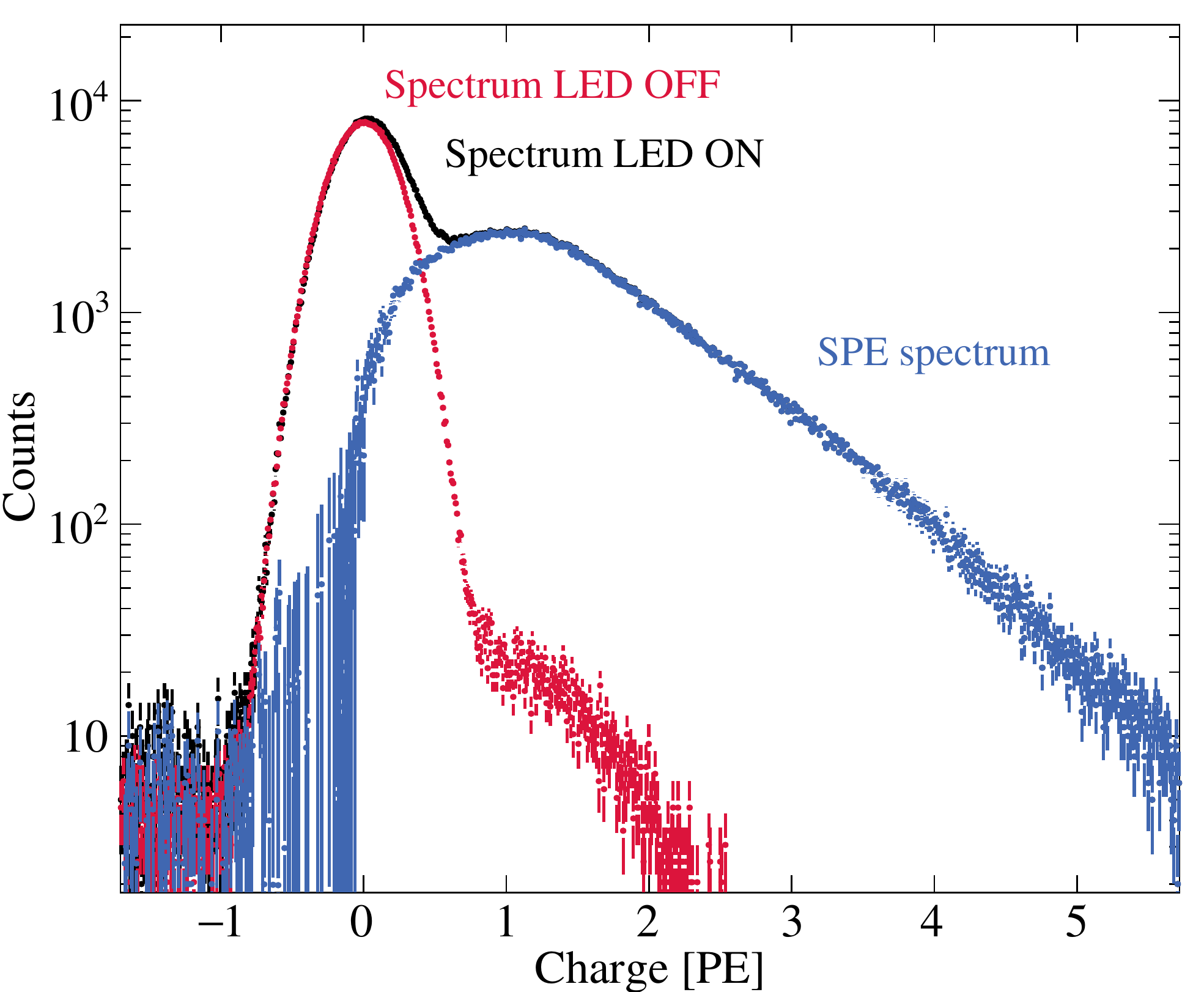}
	\caption{Gain calibration of the H10330C IR PMT. The single photo electron (SPE) spectrum is obtained by subtracting the weighted LED OFF spectrum from the LED ON spectrum}
	\label{fig:ir_spe_cal}
	\end{figure}

The response of the three photosensors is characterized via in-situ calibrations with either an ultraviolet (365\,nm) or an infrared (950\,nm) LED connected to the system via optical fibers (``LED box'' in figure\,\ref{fig:gas_system}).
The PMTs are illuminated with pulsed light signals from either of the two LEDs.
The gain is determined in a model-independent way using a subsequent blank measurement without LED light as described in~\cite{Saldanha:2016mkn}.
The UV PMTs were operated at a voltage of \SI{-1300}{V} (below the typical voltage of \SI{-1500}{V}) to avoid saturation due to large UV scintillation signals produced by the alpha interactions.
The gains of the two UV PMTs at the operating voltage were determined to be $1.6\times10^6$ and $0.8\times10^6$ for the left and right PMT, respectively. The gain of the IR-sensitive PMT was measured to be $3.52\times10^6$ at an operating voltage of $-800$\,V.
In figure\,\ref{fig:ir_spe_cal}, we show the response of the IR PMT to the LED calibration. A peak-to-valley ratio of about 1.1 is found. We also measure the dark count rate of the PMT which is $1.8\times10^5$\,Hz at the operational temperature of about $-60^{\circ}$C.

\section{Measurements and data analysis}  
\label{sec:measurements}

Before each measurement, the chamber is pumped down to a pressure of about $10^{-4}$\,mbar. Then, fresh xenon is filled into the chamber at 1\,bar, the photosensors are turned on, and the data acquisition is started. 

\begin{figure}[h]
	\centering
	\includegraphics[width=0.5\textwidth]{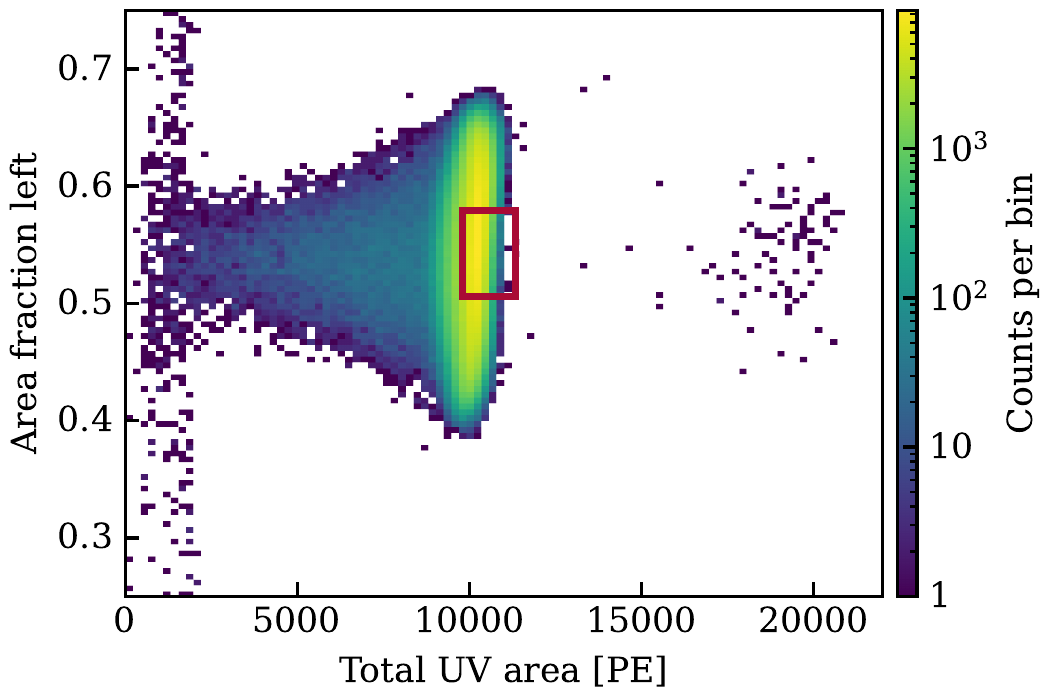}
	\caption{Area fraction of the UV PMT area seen in the left PMT as a function of the total UV area. The red box represents the region selected for the timing analysis and light yield calculation} 
	\label{fig:ana_AFL}
	\end{figure}

The events are triggered by the analog coincidence of the two UV PMTs. 
The acquired raw waveforms from each PMT are processed with the HeXe processor\,\cite{Cichon:2021phd}, which identifies peaks and computes various characteristics, including peak width and area.
To have a homogeneous alpha selection, two selection criteria are applied to the UV PMT signal: a PMT signal asymmetry cut and an energy cut. The asymmetry cut has the purpose of choosing alpha traces traveling almost exactly between the two UV PMTs.
Figure\,\ref{fig:ana_AFL} shows the fraction of UV signal in the left PMT as a function of the total UV signal size.  This parameter is defined as:
\begin{equation}
    \textrm{Area fraction left} = \frac{A_{\ell}}{A_{r} + A_{\ell}},
\end{equation}
where $A_{\ell}$ and $A_{r}$ represent the signal pulse areas in the left and right PMT, respectively.
The event distribution is centered slightly above 0.5 (half of the signal in the left PMT) and has a width corresponding to the events that leave the source under an angle smaller than 90 degrees. The shift above 0.5 is either due to a small misalignment of the source from the center of the two UV PMTs or due to uncertainties on PMT properties: PMT $QE$s or gain estimation. Asymmetry values between the 30\% and 70\% quantiles of the distribution are chosen to select alpha tracks traveling through the middle of the setup. We also make a selection on the total UV area in order to remove pile-up events (at 20\,000\,PE) and alphas with energy losses (tail toward low areas). Both cuts result in the selection of the red box in figure\,\ref{fig:ana_AFL}.

To obtain the IR time profile, we extract the arrival time of each photon in the waveform separately. The time distribution is then obtained by calculating the times of all IR photons relative to the time of the large UV signals, where the pulse time is defined as the first sample to the left of the 10\% pulse height in both cases.
For this method, we need to be able to resolve individual photons. This is feasible since the time resolution of the IR PMT is sufficiently sharp (rise- and fall-times of $\sim 1$\,ns) and the number of photons in one time window is comparably small.

We have identified a population of events characterized by low IR area and low pulse width (defined as the time duration containing 80\% of the pulse area), which we attribute to PMT noise.
To remove this noise, we apply cuts requiring each IR photon to have an area larger than 0.17\,PE and a pulse width larger than 1.85\,ns. The cut thresholds were chosen to provide a good separation between signal and noise events. To calculate the IR signal strength, we sum over all photons detected in the acquisition window of the digitizer. The infrared photons located just at the end of this window are removed to avoid a biased reconstruction.

The data presented in the following comprises two runs: one at constant pressure of 1\,bar, where the level of impurities in the gas was varied, and the other at high purity, where the pressure was varied between 500 and 1050\,mbar.
All measurements were performed at room temperature.

\section{Results and discussion}  
\label{sec:results}

This section contains the results derived from the measurements described above. It includes a study of the IR signal for different gas purity conditions, the investigation of the signal time profile, an estimation of the IR light yield, and a study of its pressure dependence.

\subsection{Effect of impurities on the IR signal}  
\label{subsec:res_purity}

\begin{figure}[h]
	\centering
	\includegraphics[width=0.5\textwidth]{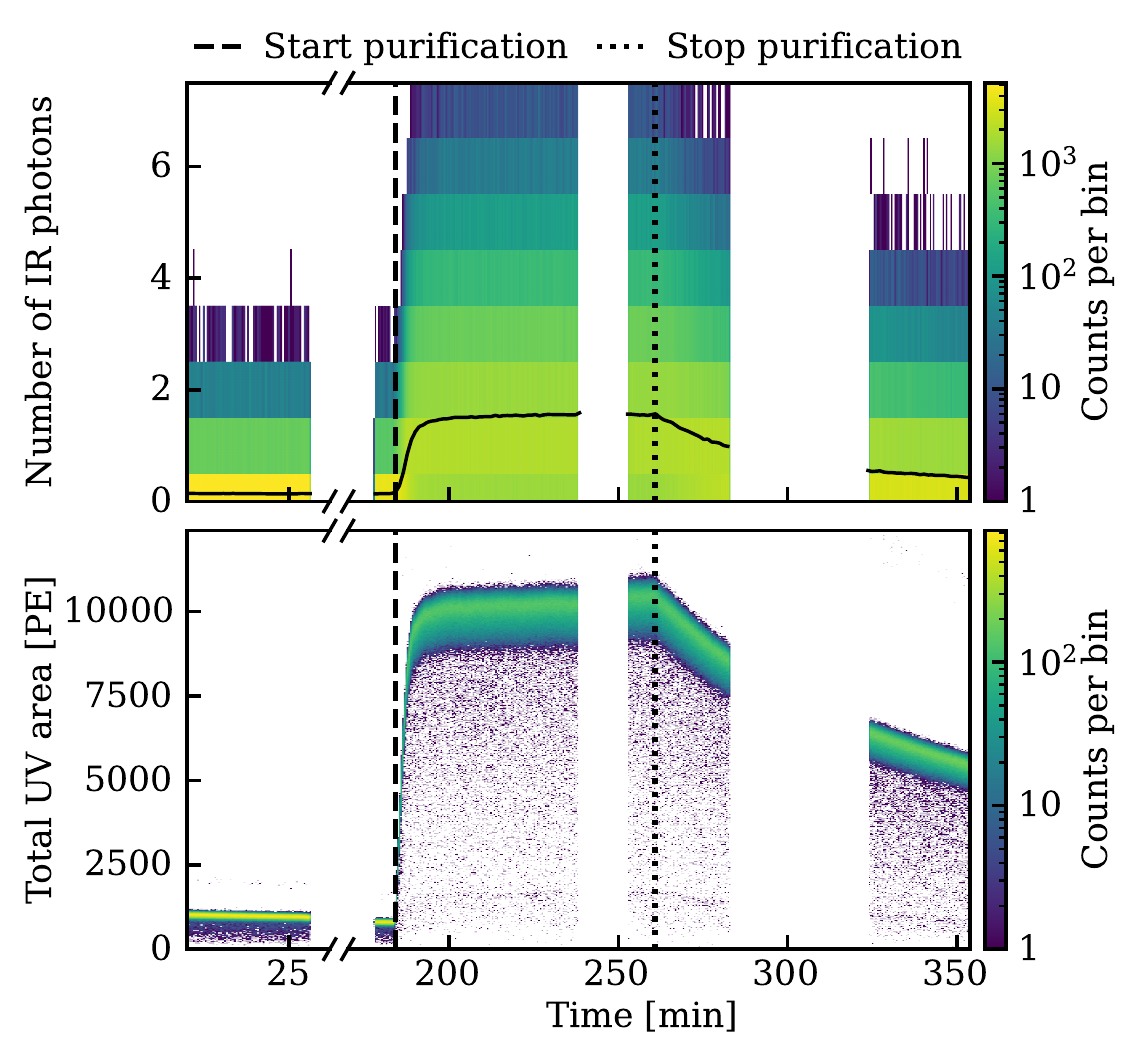}
	\caption{Two-dimensional histograms illustrating the time evolution of the number of IR photons per event (top panel) and UV  signal size (bottom panel) for different xenon purity conditions. The solid line in the top panel indicates the mean number of IR photons. The dashed line represents the time at which purification through the getter was started and the dotted line when it was stopped}
	\label{fig:res_purity}
	\end{figure}

First, we study the dependence of the infrared signal strength on the xenon purity. The chamber was filled with xenon from a storage bottle, and the measurement was started at time $t=0$\,min. Figure\,\ref{fig:res_purity} shows the signal size as a function of time, with the IR and UV signals displayed in the top and bottom panels, respectively. The two-dimensional distributions show all events after the asymmetry cut. For the IR signals, we show the total number of photons in the waveform and the black solid line indicates the mean of the distribution.
In the first measurement phase, we operated without purification.
Both the UV signal size and the mean number of IR photons decreased slightly with time in this phase, due to outgassing of impurities by the chamber and the materials inside.
The xenon purification was started by circulating it through a hot getter at about $185$\,min after the start of the run (dashed line). This was followed by a prompt increase of the signals by a factor of about 12 and 10 for IR and UV, respectively.
At time $260$\,min (dotted line), the purification was stopped by bypassing the getter, and both signals started deteriorating immediately. It can be seen that, right before the purification was stopped, the UV signal size was still slightly increasing. Therefore, somewhat higher signals can be expected for longer purification times. Although this should be a small effect, we plan to investigate it further in future measurements.

\begin{figure*}[h]
	\centering
	\includegraphics[width=\textwidth]{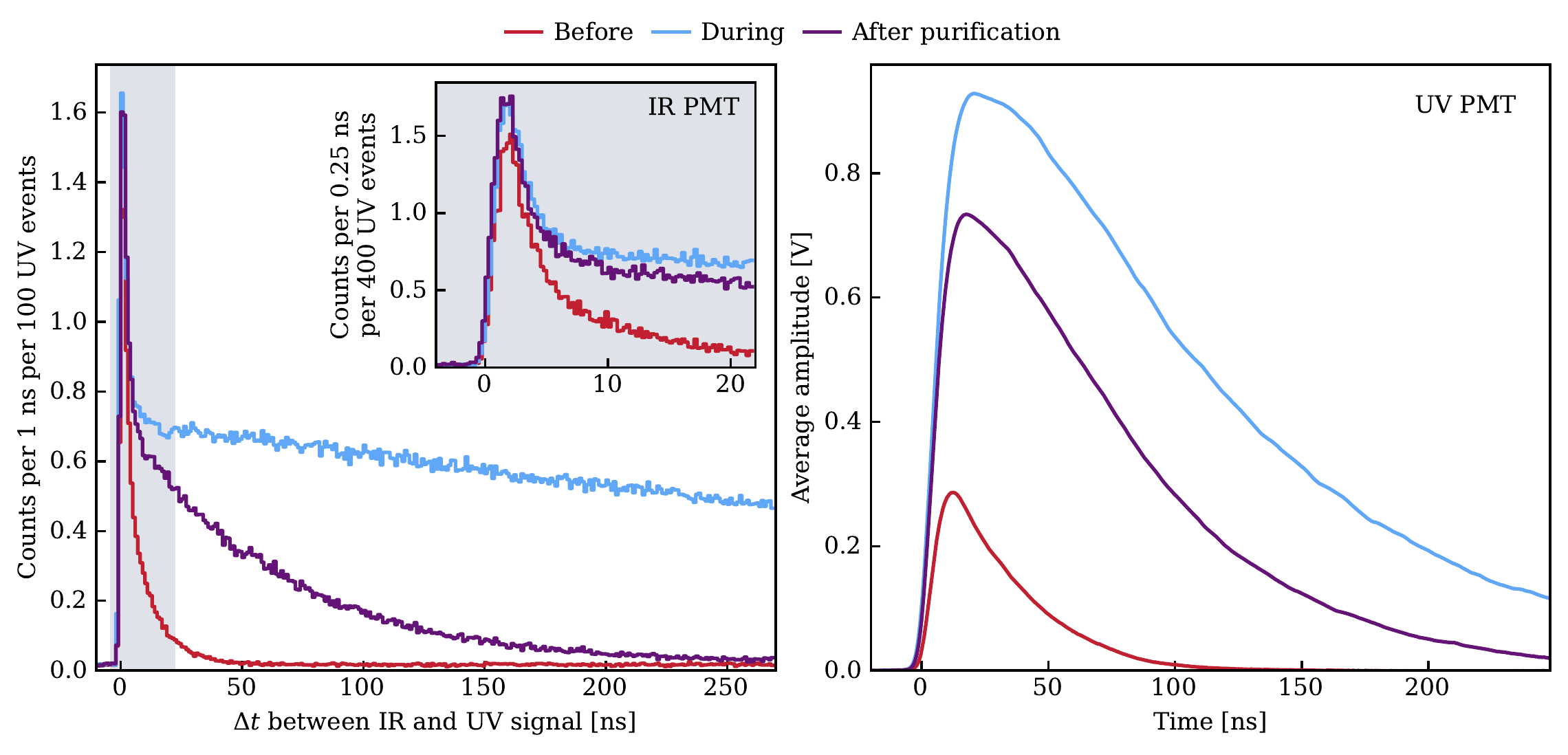}
	\caption{Scintillation response in gaseous xenon at 1\,bar for different purity levels (see regions in figure\,\ref{fig:res_purity}). Left: Arrival time of each IR photon relative to the UV time. Right: Average waveform of the UV pulses}
	\label{fig:res_timing}
	\end{figure*}

Using a photomultiplier tube as an infrared sensor allows studying the time structure of this signal with $\mathcal{O}\textrm{(ns)}$ resolution.
To investigate the time distribution of IR photons, we calculate the time of each photon as explained in section\,\ref{sec:measurements}.
We study the timing profile of the IR photons for different purity levels by choosing data in various time intervals of figure\,\ref{fig:res_purity}. Figure\,\ref{fig:res_timing} (left) shows the time distribution of IR photons.
For the time period before the purification was started (red curve), the tail of the distribution is primarily composed of a fast component with a decay time of $\sim 2$\,ns. Once the purification started (times $>200$\,min), an additional slow component with a decay time of about \SI{1}{\micro s} becomes evident (light blue). At the end of the run, the purifier is bypassed and impurities slowly accumulate again. This results in a decreasing time constant for the slow component (purple). 

For comparison, we show in figure\,\ref{fig:res_timing} (right) the average waveform of the UV pulses for the same purity conditions. 
For the period of bad purity (red), the signal is lower in amplitude and has a narrower pulse width. This is qualitatively similar to the observation in the infrared signal.

Our results show a notable increase in the intensity of the IR signal (factor 12) as the level of impurities in the xenon decreases, which is in tension with early measurements in liquid xenon\,\cite{Bressi:2000nim}. There, it was stated that the infrared signal only slightly depends on the presence of impurities.
Our hypothesis is that the impurity dependence of the IR signal arises, at least partly, from suppressed electron recombination. 
Since electrons are more likely to attach to impurities at higher impurity concentrations, they are less likely to recombine and contribute to the scintillation signal.
To a certain degree, we would expect an impurity dependence of the signal also in liquid, as electron recombination is present both in liquid and gaseous xenon.

Looking at the arrival time of IR photons, we find two time components. We attribute the fast component to excitation luminescence from both the decay of xenon excimers (D in figure\,\ref{fig:Excitons_Scheme}) or atomic transitions. The slow component could be driven by electron thermalization and recombination (B in figure\,\ref{fig:Excitons_Scheme}). We observe that the slow component is strongly affected by the purity, which could be interpreted as a reduction of recombination electrons due to attachment to electronegative impurities, such as O$_2$.
We can exclude that secondary scintillation by the drifting electrons is a major contribution to the observed signal. The electric field in the analysis volume is below 400\,V/cm and, therefore, below the $\sim 1$\,kV/cm$\cdot$bar required for this process (see\,\cite{Feio_1982} and references therein). Below this threshold, neutral bremsstrahlung emission has been observed\,\cite{henriques2022neutral}, which might contribute to the observed slow component.
The fast component is almost not affected by the change in impurity concentration.
Similarly, we find that the size and shape of the UV pulse vary with purity, becoming smaller and narrower for decreasing purity. This is likely the result of absorption of UV light via impurities\,\cite{Ozone2005} as well as quenching processes affecting the slow scintillation decay component (triplet), which has a lifetime of $\sim 100$\,ns\,\cite{Murayama:2014wwa}. Furthermore, the absence of recombination signal at delayed times can be partially responsible for the narrower distribution, as electrons are attached to impurities.

Following our hypothesis that the slow IR component primarily originates from recombination luminescence, we can estimate the amount of photons arising from direct excitation relative to the ones coming from recombination. 
Note that, since our digitizer has a small acquisition window compared to the length of the pulse, we need to extrapolate and add the missing part of the slow component  (see section\,\ref{subsec:res_ly}).
We find that the ratio of excitation luminescence to recombination luminescence in the IR would be below 1\%.

\subsection{Estimation of the infrared light yield}  
\label{subsec:res_ly}

The light yield, defined as the number of photons emitted per unit energy deposited, can be expressed as:
\begin{equation}
    Y = \frac{\mu}{E_{\alpha} \cdot \epsilon \cdot QE},
\end{equation}
where $\mu$ is the number of detected photons per event, $E_\alpha$ is the deposited energy, $QE$ is the PMT quantum efficiency, and $\epsilon$ is the solid angle acceptance.

The number of detected photons per event, $\mu$, is estimated via the mean value $\tilde{\mu}$.
For the infrared signal, $\tilde{\mu}$ is the mean number of observed photons per event. For the UV spectrum, $\tilde{\mu}$ is the mean of the summed UV area after applying a strict energy cut (as shown in figure\,\ref{fig:ana_AFL}) such that alphas with losses in the source or the aperture are not considered. 
Due to the short acquisition window of the digitizer (total of 320\,ns), the number of registered IR photons needs to be corrected for the amount of missing signal to obtain $\mu$. To estimate this, we fit a function to the tail of the IR photon time distribution. Since the underlying distribution of the slow component is unknown, we use three simple fitting functions: linear, exponential, and a function proportional to $1/(1+\Delta t / T_\mathrm{r})^2$ with the recombination time $T_\mathrm{r}$, as motivated by the recombination process in~\cite{Kubota:1979xx}.
The computed light yields are scaled based on the extrapolated fraction of the signal outside the acquisition window, which varies between different fitting models. For 1\,bar xenon pressure, the linear fit ($lin$) yields a correction factor of about 2, while the exponential fit ($exp$) yields about 3.5, and the recombination model fit ($rec$) yields about 6.5.

The solid angle acceptance is estimated via a simplified Monte Carlo simulation. In the simulation, photons are emitted isotropically from the entire alpha track and the intersections with the UV PMT photo-cathode or the lens of the IR PMT are recorded. The acceptance for 1\,bar xenon pressure is 15.5\% for each UV PMT and 0.25\% for the IR PMT. This approach neglects possible additional signals due to reflections on the surrounding surfaces, constituting only an approximate estimate of the respective solid angle acceptances. Note, however, that the inside of the chamber is not polished, likely making it a bad reflector. In addition, quartz (the material of the UV PMT windows) transmits about 90\% of the infrared light and can thus also be considered a bad reflector.
The length of the alpha track is calculated separately using the SRIM simulation toolkit\,\cite{Ziegler:2010}.
The calculation of the deposited energy takes into account the small fraction of energy that is lost in the region between the source surface and the aperture in front of it. For 1\,bar pressure the visible energy of the alpha corresponds to 5.1\,MeV which is about 92\% of the total alpha energy.
Finally, the quantum efficiency of the sensors is considered: $(35\pm3)$\% for the UV PMTs and $(9.0\pm0.5)$\% for the IR PMT. 

The measurement of the IR light yields is subject to a range of systematic and statistical uncertainties. Systematic uncertainties include those related to the gain of the PMT, energy deposition before the aperture, solid angle acceptance, and PMT $QE$. The relative PMT $QE$ uncertainty, at $5.6\%$, is the largest of these. However, the dominant systematic uncertainty is due to the waveform correction, resulting from the short acquisition window of the digitizer as described above. For this reason, we give the results for all three signal extrapolations.
Statistical uncertainties arise from the estimation of the mean number of PE and the fit to the slow component.

In this study, our focus is on investigating the IR scintillation response, and we study the UV signal just to validate our procedure.
Thus, we only give the signal size in the observed window as a lower value.

To calculate the light yields, we use a dedicated measurement taken over 30\,min and with a constant high purity.
The light yield results for the IR and UV signals of gaseous xenon at 1\,bar and room temperature are reported in table\,\ref{tab:table-yields}.

\begin{table}[h]
\centering
\caption{\label{tab:table-yields}IR and UV light yields for gaseous xenon at 1\,bar pressure and room temperature. The IR light yield is given for three different extrapolations of the signal beyond the acquisition window}
\begin{tabular}{ lll}
 \toprule
    {\bf Signal} & {\bf Ph. det.} $\tilde{\mu}$ {\bf [PE]} &{\bf Yield $Y$ [ph/MeV]} \\
 \midrule 
\multicolumn{1}{l}{IR ($lin$)} &  \multirow{3}{*}{$\Biggl\}$$1.7456 \pm 0.0026$}& \multicolumn{1}{l}{$3620 \pm 50\mathrm{_{\,stat}} \pm 230\mathrm{_{\,syst}}$} \\
\multicolumn{1}{l}{IR ($exp$)}&  & \multicolumn{1}{l}{$6120 \pm 110\mathrm{_{\,stat}} \pm 400\mathrm{_{\,syst}}$} \\
\multicolumn{1}{l}{IR ($rec$)}&  & \multicolumn{1}{l}{$11150 \pm 220\mathrm{_{\,stat}} \pm 700\mathrm{_{\,syst}}$} \\[0.4em]              UV & $10155.0 \pm 0.4$ & $>18700$  \\
 \bottomrule          
 \end{tabular}
\end{table}

The yield of the IR scintillation light is in the same order of magnitude as that of UV even though the number of IR photons detected per event is low compared to the UV.
This is due to the very small solid angle for detection and the low $QE$ of the IR-PMT (see section\,\ref{sec:measurements}). In addition, a larger fraction of the IR signal is outside the acquisition window.
Our reported IR light yield values may underestimate the actual yield as our PMT covers only part of the IR emission spectrum (see figure\,\ref{fig:ir_sensitivity}). Although the PMT's spectral sensitivity window covers the entire continuous excimer emission, there may be additional atomic transitions outside of this range that could contribute to the overall yield.

The infrared light yield values are all lower than the value of 20\,000\,ph/MeV measured in~\cite{Belogurov:2000}. The reason for the disagreement is currently under investigation.
One difference between the measurements is the xenon pressure, 2\,bar in~\cite{Belogurov:2000} and 1\,bar in our case. We observe a small dependence of the signal with pressure (see section\,\ref{subsec:ir_pressure}), but we have no measurements above 1.05\,bar and therefore cannot judge if this is the reason.
Another difference is that our setup does not have, at the moment, a homogeneous electric field along the particle track. An upgrade of the system is planned to repeat the measurements at well-defined field values.
Finally, we note that the spectral response of the infrared sensors is different in both setups. While ours is sensitive from $(950-1650)$\,nm, the one used in~\cite{Belogurov:2000} is sensitive from $(700-1600)$\,nm. We expect a large fraction of the IR signal to come from the excimer continuum at $\sim$ \SI{1.3}{\micro m}. Nevertheless, we cannot rule out the possibility that unknown strong lines with wavelengths below 800\,nm could account for part of the observed difference.

For the UV component, the average energy required to create a UV photon in xenon for an alpha particle is reported to be $(35-50)$\,eV\,\cite{Leardini:2021qnf,Saito:2003xxx}. This would correspond to a light yield $\sim (20\,000-28\,000)$\,ph/MeV, which is in the same order of magnitude as our measurement at $\sim 18\,700$\,ph/MeV.

\subsection{Pressure dependence of the IR signal}
\label{subsec:ir_pressure}

\begin{figure}[h]
	\centering
    \includegraphics[width=0.48\textwidth]{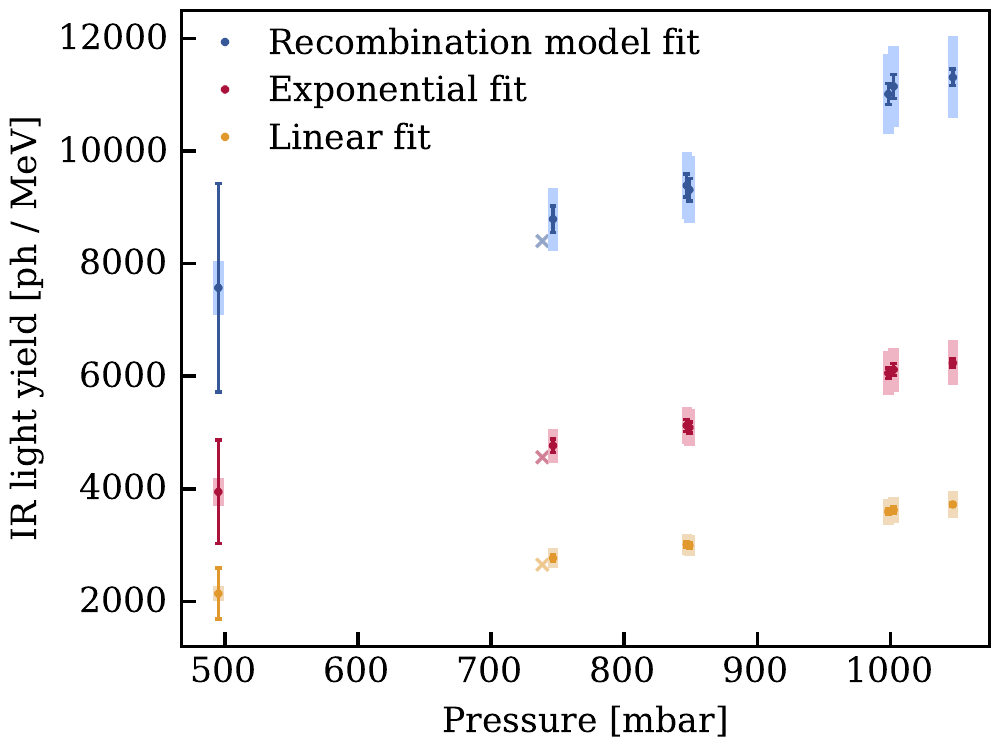}
	\caption{IR light yield as a function of pressure for three different extrapolation methods. Uncertainties are shown as error bars (statistical uncertainty) and shaded bands (systematic uncertainty). Note that one measurement (transparent crosses) was excluded from the analysis due to an ongoing increase in xenon purity during the corresponding data-taking period}
	\label{fig:press_study}
	\end{figure}

We also study the dependence of the light yield on the xenon gas pressure. For this purpose, we vary the gas pressure in the range of 500 to 1050\,mbar and determine the light yield with the same procedure as described above. Figure\,\ref{fig:press_study} shows the IR light yields as a function of pressure for the three signal extrapolation models described in section\,\ref{subsec:res_ly}.
The data presented in the plot were acquired over about 5.5 hours, with a 30-minute measurement for each data point. We started the run at a pressure of about 750\,mbar and gradually increased up to about 1050\,mbar (two consecutive 30\,min acquisitions, combined in the plot), before decreasing the pressure again with the same steps. In this way, there are two measurements for each pressure, with a time difference in between. The first data point at 750\,mbar is slightly lower than the second one because the purifier was only running since 10\,min and the signal was still slightly growing.
Thus, we exclude this measurement from the analysis and only show it in the plot as transparent crosses for reference.
At the end of the run, the pressure was lowered to 500\,mbar. However, due to the low pressure, the UV PMTs experienced a high voltage breakdown after about five minutes. As a result, the uncertainties of the data points at 500\,mbar are statistics dominated.

 All values have been corrected to account for the solid angle acceptance of the IR PMT due to the longer alpha track at low pressures. Indeed, the average length of the alpha track (calculated by SRIM) varies from 2.3\,cm at 1050\,mbar to 4.8\,cm at 500\,mbar. Thus, the collection efficiency increases for decreasing pressures from 0.22\% at 1050\,mbar to 0.30\% at 500\,mbar. We also correct the alpha energy deposition, since the fraction of energy deposited after the aperture is pressure dependent as well. The fraction of the alpha energy observed after the aperture changes from 91.7\% at 1050\,mbar to 96.9\% at 500\,mbar.

For the pressure range studied here, we observe an increase in the light yield with pressure, which is noticeable across all signal extrapolation models. 
At low pressures, the average distance between the freed electrons and the ions is larger, increasing the probability of the electrons escaping recombination\,\cite{Mimura:2010xx}. Additionally, due to the presence of a non-negligible electric field in the gaseous volume, we anticipate that at lower pressures, an even greater number of electrons will be drifted away, leading to a slightly reduced signal from recombination (see similar effect in the UV\,\cite{Suzuki:1982yy}). 
The change in pressure also leads to a shift in the emission spectrum, as pointed out in~\cite{Borghesani:2001xx,Borghesani_2007,Borghesani:2008xx}. The range tested here is, however, small enough that the complete excimer spectrum is contained in the sensitive region of the IR-PMT and a minor effect is expected.

\begin{figure*}[ht]
	\centering
	\includegraphics[width=\textwidth]{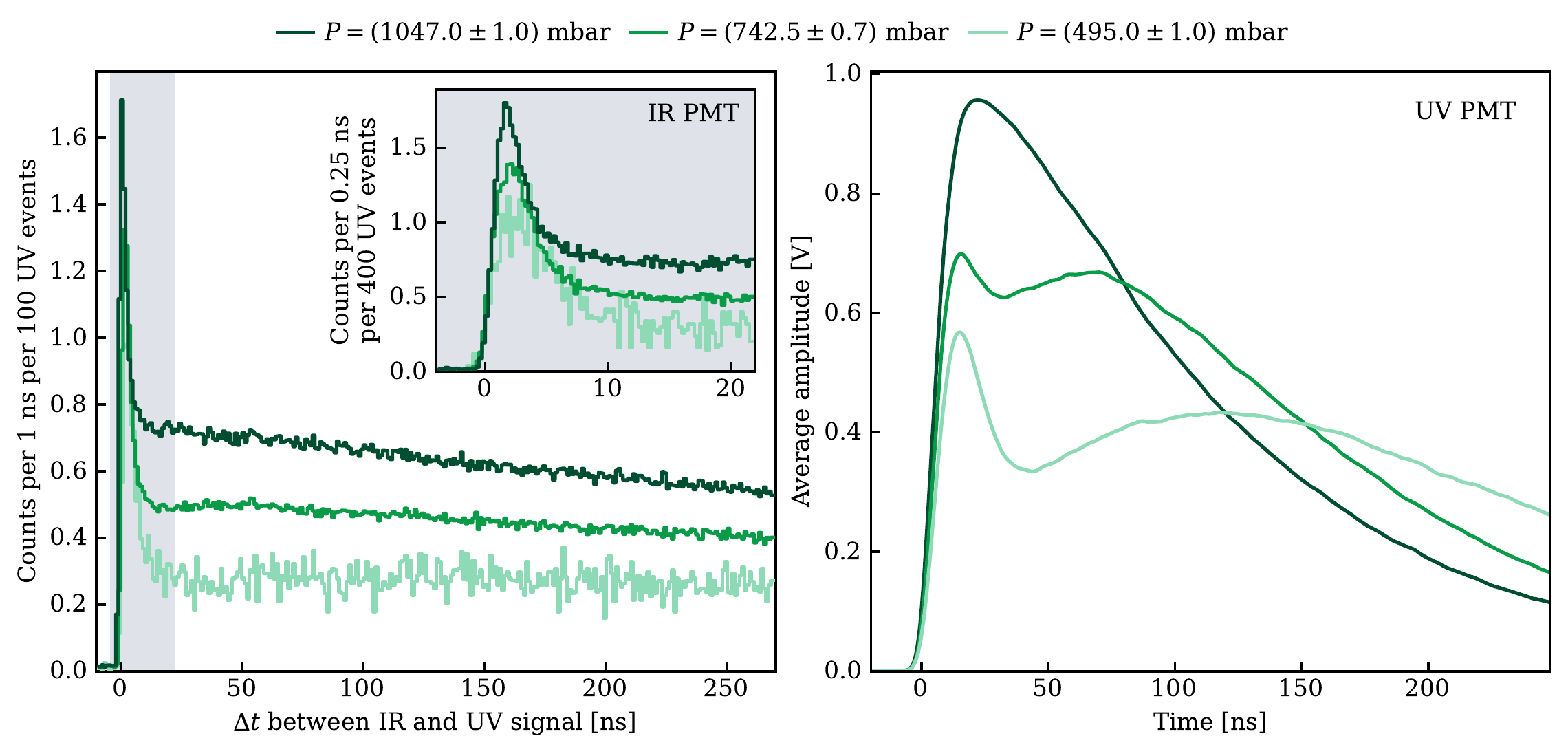}
	\caption{Left: Arrival time of IR photons (like figure\,\ref{fig:res_timing} left) for different xenon gas pressures. At the lowest pressure, we acquired data for only about five minutes, whereas for the other two pressures, one hour of data is used. Right: Average waveform of the UV pulses for gaseous xenon for different pressures}
	\label{fig:press_timing}
	\end{figure*}

Figure\,\ref{fig:press_timing} (left) shows the arrival time of the IR photons for three of the pressure values studied.
We observe that the slow component is slightly suppressed for lower pressure and that the time constant is longer. This is in agreement with our hypothesis that recombination probability decreases with decreasing pressures. 

The UV pulses show for low pressures a double bump shape (see figure\,\ref{fig:press_timing} right). This behavior is also observed in the data of\,\cite{Bolotnikov:1999,Ribitzki:1994,Takahashi:1982xx}. 
As in these early publications, we would interpret the delayed bump (``xenon afterglow") in the time spectrum as originating from electron recombination processes. For decreasing pressure values, the recombination time increases and the second bump is further delayed. 
Other processes that could play a role in the UV time structure are the excimer formation, whose time should become longer for low pressures, or a contribution from the 3\textsuperscript{rd} continuum (decay of Xe$_2^{2+}$, for instance, at 270\,nm)\,\cite{Leardini:2021qnf,Millet:1978}. Future measurements are planned to investigate this further.
For the lower pressure time profiles in figure\,\ref{fig:press_timing} (left) one can also see a slight indication of a dip for arrival times between $\sim$10 and 50\,ns. This feature might be an effect similar to the double bump structure in the UV signal (see figure\,\ref{fig:press_timing} right).

\section{Discussion and outlook}  
\label{sec:discussion}

Our measurements confirm the emission of infrared light by gaseous xenon as measured in earlier works\,\cite{Carungo:1998xx,Belogurov:2000,Borghesani:2001xx,Borghesani:2008xx}.
We find that the IR signal is strongly affected by electronegative impurities, in contrast to early measurements in liquid xenon\,\cite{Bressi:2000nim}.
Looking at the time structure of the signal, we observe two main components. Although we do not have a consistent model describing all our data, we think that those components likely originate from direct excitation (fast component) and from electron recombination (slow). We show that the time constant of the slow component decreases with an increasing impurity concentration in the xenon, while the fast component stays almost constant.
Two distinctive ionization regions for energy depositions from alpha particles are reported in literature: a dense core and a penumbra around it. Accordingly, recombination processes have two different components: a fast one for the core and a slow one for the penumbra\,\cite{Mimura:2010xx}. 
While our data does not provide conclusive evidence, it is possible that the slow component originates from the core region. The electrons in the penumbra are likely drifted away by the electric field in the chamber.
Independent of the extrapolation method used to calculate the IR light yield, our measurements indicate that the IR light yield is lower than that of the UV at a gas pressure of 1\,bar. This might be related to the different population of states in direct excitation and recombination. As the energy deposition of charged particles populates all xenon excited states, some states might decay without the emission of IR photons. Since recombined electrons populate higher lying excited states, we expect that their decay emit both IR and UV photons.
Furthermore, some of the de-excitation paths through IR atomic levels are outside the sensitivity of our sensor. Our data additionally shows that the light yield increases with pressure between \SI{500}{mbar} and \SI{1050}{mbar}.

While the hypothesis that the majority of the infrared signal (slow component) comes from electron recombination is supported by all IR measurements, this interpretation does not align with the observed shape of the UV pulses, which lack a long-lived component. Additionally, we find that the fraction of the fast IR component is consistently below 1\% in all our measurements, which differs from the exciton-to-ion ratio obtained via the UV component in the literature. The UV ratio approaches a value around $(60-65)$\% for decreasing pressure values\,\cite{Saito:2003xxx,Bolotnikov:1999}.
The difference could be partly explained by strong atomic lines outside the sensitivity window of our PMT in the case of direct excitation.

Measurements of the infrared signal with a homogeneous electric field (zero-field and higher electric field) are planned for the upcoming runs with an upgraded setup. 
In the future, we plan to investigate the dependence of the infrared emission in gaseous xenon on the electric field strength and particle-type, similar to our study of UV yields in liquid xenon\,\cite{Jorg:2021hzu}. We also aim to acquire data with a longer time window to alleviate the dominant systematic uncertainty and, thus, allow for more precise quantification of the IR light yield. 
These improvements are expected to provide a better understanding of the origin of the IR time components and enable a precise modeling of the underlying processes\,\cite{Azevedo:2017egv}.
The goal of our measurements is to explore the potential of infrared radiation to advance current astroparticle physics detectors. Ultimately, we aim to investigate if this can improve energy resolution and/or particle separation in double-phase liquid xenon detectors\,\cite{Aalbers:2016jon,Aalbers:2022dzr}.

\vspace{0.6cm}
\noindent
\textbf{Acknowledgments} We thank Andreas Ulrich, Edgar S\'anchez Garc\'ia and Diego Gonz\'alez D\'iaz  for the valuable discussions regarding the scintillation process in noble gases.
We thank our technicians Steffen Form, Michael Reißfelder, and Hannes Bonet for their technical support during the construction of the system. We acknowledge the support of the Max Planck Society.

\vspace{0.6cm}
\noindent
\textbf{Data Availability Statement} The infrared data is available online at \url{https://www.doi.org/10.5281/zenodo.7936373}.

\vspace{3cm}
\bibliographystyle{utphys}
\bibliography{IR_GXe}   

\end{document}